\documentstyle[aps,prd,epsfig]{revtex}
\begin{document}
\title{Possible $Z$-width probe of a ``brane-world" scenario
for neutrino masses}\vfill
\author{S. {\sc Moussa}$^{(a)}$\footnote{Electronic address:
{\tt sherif@phy.syr.edu} }\quad S. {\sc
Nasri}$^{(a)}$\footnote{Electronic address: {\tt
snasri@phy.syr.edu}} \quad F. {\sc Sannino}$^{(b)}$
\footnote{Electronic address: {\tt francesco.sannino@nbi.dk}}
\quad J. {\sc Schechter}$^{(a)}$ \footnote{Electronic address:
{\tt schechte@phy.syr.edu}}} \vfill
\address{$^{(a)}$ Department of Physics, Syracuse University,
Syracuse, NY 13244-1130, USA.~\\ {\rm $^{(b)}$ NORDITA},
Blegdamsvej 17, DK-2100 Copenhagen \O, Denmark.}
\date{August 2001}
\maketitle

\begin{abstract}
The possibility that the accurately known value of the Z width
might furnish information about the coupling of two neutrinos to
the Majoron (Nambu-Goldstone boson of spontaneous lepton number
violation) is proposed and investigated in detail. Both the
``ordinary" case and the case in which one adopts a ``brane" world
picture with the Majoron free to travel in extra dimensions are
studied. Bounds on the dimensionless coupling constants are
obtained, allowing for any number of extra dimensions and any
intrinsic mass scale. These bounds may be applied to a variety of
different Majoron models. If a technically natural see-saw model
is adopted, the predicted coupling constants are far below these
upper bounds. In addition, for this natural model, the effect of
extra dimensions is to decrease the predicted partial Z width, the
increase due to many Kaluza-Klein excitations being compensated by
the decrease of their common coupling constant.
\end{abstract}
\begin{flushleft}
\footnotesize
~~~~~~~~~~~~~~~~SU--4240--747, NORDITA-2001-23 HE
\end{flushleft}

\section{Introduction}

\label{uno}

At present, there are many hints that our working models of
elementary particle interactions and gravity may be unified at a
short distance scale by a string-type theory. The practical
implementation of such ideas is conceivably ``around the corner"
but also probably very far off. In such a situation it makes sense
to study simpler models which display some features of the string
theory. Especially interesting is the (Kaluza-Klein) idea of extra
dimensions and the possibility that we live on a ``brane" immersed
in these extra dimensions \cite{1,RS}. It is also natural to expect that
such
models \cite{DDG,ADDM,DS,BGS,IV} could shed some light on the theory of
massive neutrinos
which seem to be reluctant participants in the so called
``standard-model". A popular assumption, which is reasonable to
study in detail, is the requirement that particles carrying
non-trivial $SU(3)_c\times SU(2)_L\times U(1)_Y$ quantum numbers
and the associated gauge fields are confined to the brane while
others (like the graviton) may propagate also in the extra dimensions.
Many authors have thus investigated the possibility that right
handed neutrinos might propagate in the extra dimensions and that
their resultant suppressed couplings to the usual left handed
neutrinos could account for the low scale of neutrino
masses\cite{DDG,ADDM}. A large number of authors \cite{FP,MN,DK,C,BDN}
have studied the
constraints on such models due to experiment and observation. An
alternate approach\cite{MPP}, corresponding to the conventional see-saw
mechanism, has also been investigated by some authors. In this
approach a Higgs singlet, which carries no standard model gauge
quantum numbers, is allowed to propagate in the extra dimensions
and give mass to right handed neutrinos which, for simplicity, are
assumed to live on the brane. This model does not automatically result in
the
correct neutrino mass scale, but may do so in more sophisticated
versions. In any event it is interesting to study a version in
which the lepton number is spontaneously broken so that a
Goldstone boson called the ``Majoron" is present. (See also \cite{MRS})

In this paper we shall study a simple model of this type and
calculate the decay rate for the intermediate vector boson $Z$ to
go to two neutrinos and the Majoron (denoted by $J$) or one of its
``Kaluza-Klein" excitations. Depending on the exact nature of the
extra dimension scenario, there may be many such excitations so
that the lepton number violating process $Z\rightarrow \nu \nu
(J)$ might be expected to be large enough to, some day, be detected by a
subtraction measurement. According to the latest ``Review of
Particle Physics"\cite{rpp} the $Z$ width is:
\begin{equation}
\Gamma(Z)=2.4952\pm 0.0022~{\rm GeV} \ ,
\label{zwidth}
\end{equation}
and the ``invisible" partial width is
\begin{equation}
\Gamma({\rm invisible})=499.4\pm 1.7~{\rm MeV} \ .
\label{invisiblewidth}
\end{equation}
The uncertainties in these expressions give an idea of the partial
width needed for possible detection. The related neutrinoless
double beta decay process $n+n\rightarrow p+p+e^{-}+e^{-}+(J)$ has
already been treated in a model of the present type \cite{MPP}. A detailed
discussion of supernova constraints in the 3 $+$ 1 dimensional theory
has very recently been given in \cite{TPV}.


Section II contains a brief review of the Majoron model and also
its extension to the extra dimensional brane-world picture. In
section III the partial width for the lepton number violating
process Z $\rightarrow $ two neutrinos plus a particular majoron
excitation is expressed as an integral over phase space. The
integrand is evaluated to leading order in the neutrino mass,
$m_\nu$.In this simplifying limit, an overall factor $m_\nu^2$
carries its $m_\nu$ dependence. The integral itself is evaluated
in section IV for this process in the usual 3 $+$ 1 dimensional
theory with a single zero mass majoron. This is a little delicate
since the main contribution arises from near the phase space
boundary, just outside of which lurks a singularity. It proves
instructive to evaluate the integral analytically. Section V
contains the calculation for the partial width of Z decay to two
neutrinos plus a particular majoron excitation in the extra
dimensional theory. An analytic approximation of the numerically
obtained rate integral, based on the approach of the previous
section, permits convenient integration over the majoron tower in
the general case. Finally, a brief summary is presented in section
VI.

 \section{Majoron model}

First, we briefly review the original Majoron model of Chikashige,
Mohapatra and Peccei \cite{CMP}. It is a model for generating
spontaneously the broken lepton number associated with massive
Majorana neutrinos. Here, the notations of \cite{SV} will be followed. In
addition to the usual Higgs doublet
\begin{equation}
\phi =  \left(
\begin{array}{c}
\phi ^{+} \\ \phi ^{0}
\end{array}
\right) \qquad l=0
\label{usualhiggs}
\end{equation}
which has lepton number $l$ equal to zero, the model contains an
electrically neutral complex singlet field
\begin{equation}
\Phi \qquad l=-2 \ .
\label{singlethiggs}
\end{equation}
The kinetic terms of the Lagrangian are:
\begin{equation}
-\frac{1}{2} \partial_{\mu}\phi^{\dagger}\partial_{\mu}\phi
-\frac{1}{2}\partial_{\mu}\Phi^{\ast}\partial_{\mu}\Phi \ .
\label{usualkin}
\end{equation}
(The factor $1/2$ is a convenient convention and we also use the
Minkowski metric convention $(x_0=it,x_1,x_2,x_3)$.) It is
required that the Higgs potential constructed from $\Phi$ and
$\phi$ conserves lepton number. The vacuum values are:
\begin{eqnarray}
<\Phi > &=&<\Phi^{\ast}>=X\ ,  \nonumber \\
<\phi ^{0}> &=&<{\phi^{0}}^{\dagger}>=\lambda
\approx 2^{-\frac{1}{4}}G_{F}^{-\frac{1}{2}}\
,
\label{vacvalues}
\end{eqnarray}
where $G_F$ is the Fermi constant and $X$ (whose non-zero value
violates lepton number) sets a new scale in the theory. We assume
the theory contains three two component neutrino spinors $\rho_1,
\rho_2, \rho_3$ with $l=1$, belonging to $SU(2)_L$ doublets and
three two component spinors $\rho_4, \rho_5, \rho_6$ with $l=-1$
which are singlets under $SU(2)_L\times U(1)_Y$. These are united
as
\begin{equation}
\rho^T=\left(\rho_1^T, \rho_2^T, \rho_3^T,\rho_4^T, \rho_5^T,
\rho_6^T\right) \ ;
\label{sixnus}
\end{equation}
all the $\rho_a$ have the same Lorentz transformation property.
Then the (lepton number conserving) Yukawa terms involving the
neutrinos may be written as:
\begin{equation}
{\cal L}_{yukawa}=-\frac{1}{2} \rho^T \sigma_{2} \left(
\begin{array}{cc}
0 &\frac{\phi^0}{\lambda}{\cal D} \\ \frac{\phi^0}{\lambda}{\cal
D}^T & \frac{\Phi^{\ast}}{X}{\cal M}_H
\end{array}
\right) \rho + {\rm h.c.}\ ,\label{yukawa}
\end{equation}
where $\sigma_2$ is the Pauli matrix. The $3\times 3$ matrix
$\frac{1}{\lambda}{\cal D}$ represents the ``Dirac-type" coupling
constants for the bare light neutrinos while the $3\times 3$
matrix $\frac{1}{X}{\cal M}_{H}$ represents the Majorana type
coupling constants for the bare heavy (or ``right handed")
neutrinos. As a whole Eq.~(\ref{yukawa}) is just a generic
``see-saw mechanism" \cite{ss}. It is necessary to diagonalize the matrix
by
a unitary transformation:
\begin{equation}
\rho = U \nu \ ,
\label{Utransf}
\end{equation}
to physical fields $\nu$. This can be carried out \cite{SV}
approximately as a power series expansion in
\begin{equation}
\epsilon = {\cal O} \left( \frac{\cal D}{{\cal M}_H}\right) \ .
\label{eps}
\end{equation}
We will focus attention on the three light physical neutrinos
$\nu_{1},\nu_2,\nu_3$. These will acquire Majorana masses
$m_1,m_2,m_3$ which are of order $\epsilon^2 {\cal M}_{H}$ (which
is just the counting of the see-saw mechanism). For our present
purpose we need the coupling of the Majoron $J$, identified as
$J={\rm Im}\Phi$, to the physical neutrino fields
$\nu_1,\nu_2,\nu_3$:
\begin{equation}
{\cal L}_{J}=i\frac{J}{2}\sum_{a,b=1}^3 \nu^T_{a}\sigma_2 g_{ab}
\nu_b + {\rm h.c.} \ .
\label{lj1}
\end{equation}
It turns out \cite{SV} that the coupling constants have the
expansion:
\begin{equation}
g_{ab}=-\frac{1}{X}m_a \delta_{ab} + {\cal O}\left(\epsilon^4
{\cal M}_{H} \right) \ ,
\label{gexp}
\end{equation}
where\footnote{Note that the expression for the $\epsilon^4 {\cal
M}_H$ terms given in Eq.~(6.8) of \cite{SV} should be
symmetrized.} the leading term is seen to be diagonal in
generation space. Rewriting, for convenience, this leading term
using four component ordinary Dirac spinors
\begin{equation}\psi_{a} =\left(
\begin{array}{c}
\nu _{a } \\ 0\end{array} \right) \ ,
\label{4comp}
\end{equation}
in a $\gamma_5$ diagonal representation of the Dirac matrices, we
get:
\begin{equation}
{\cal L}_{J}=i\frac{J}{2X}\sum_{a=1}^3 m_a \left(\psi^T_a
C^{-1}\frac{1+\gamma_5}{2}\psi_a +
\bar{\psi}_a\frac{1-\gamma_5}{2}C \bar{\psi}^T_a\right) \ .
\label{lj2}
\end{equation}
Here $C$ is the charge conjugation matrix of the Dirac theory.

Now let us consider how this treatment gets modified when we allow
the singlet field $\Phi$ to propagate in $\delta$ extra spatial
dimensions. These extra dimensions, denoted as $y_{i}$ with
$i=1,\ldots \delta$, will be assumed to be toroidally compactified via
the identification $y_i=0$ and $y_i=2\pi R_i$. {}For simplicity
all $R_i$ will be taken equal to the same value $R$. It is
convenient to take $\Phi\left(x,y\right)$ to continue to carry the
``engineering dimension" one as it would in $3+1$ dimensional
space-time. Then the kinetic term of the dimensionless action in
$(3+\delta)+1$ space-time
\begin{equation}
S=-\frac{1}{2}\int d^4x d^{\delta}y\, M^{\delta}_S \,
\partial_{\mu}\Phi^{\ast} \partial_{\mu}\Phi
 \ ,
\label{Skin}
 \end{equation}
 includes a rescaling mass $M_S$ which represents the intrinsic
 scale of the resulting theory. With a Fourier expansion with
 respect to the compactified coordinates
\begin{equation}
\Phi\left(x,y\right)={\rm Norm} \sum_{n_1,n_2,\ldots,n_{\delta}}
\Phi_{n_1,n_2,\ldots,
n_{\delta}}\left(x\right)\exp\left[\frac{i}{R}\left(n_1 y_1+n_2
y_2+ \cdots \right)\right] \ , \label{normalization}
\end{equation}
up to an additive constant
where ${\rm Norm}=\left[2\pi M_S R\right]^{-\frac{\delta}{2}}$,
the kinetic action reads
\begin{equation}
S=-\frac{1}{2}\int d^4x \,
\left[\partial_{\mu}\Phi^{\ast}_{n_1,n_2,\ldots}(x)
\partial_{\mu}\Phi_{n_1,n_2,\ldots}(x)+\frac{1}{R^2}\left(n_1^2+n_2^2+ \cdots
\right)\Phi_{n_1,n_2,\dots}^{\ast}(x)\Phi_{n_1,n_2,\ldots}\right]
\ .
\label{kinaction}
\end{equation}
This expression with \begin{equation} \Phi_{n_1,n_2,\ldots}(x) = {\rm
Re}\Phi_{n_1,n_2,\ldots,}(x)+ i\, J_{n_1,n_2,\ldots}(x) \ ,
\label{complex}
\end{equation}
shows that each Kaluza-Klein (i.e. Fourier component) field
receives a mass squared increment
\begin{equation}
\Delta m^2_{n_1,n_2,\ldots}=\frac{1}{R^2}\left(n_1^2+n_2^2+\cdots
n_{\delta}^2 \right) \ . \label{increment}
\end{equation}
The true, zero-mass, Majoron is $J_{0,0,\ldots,0}(x)$ and will
receive no other mass squared increment. However the fields ${\rm
Re} \Phi_{n_1,n_2,\ldots}(x)$ will receive a substantial increment
from the pure Higgs sector of the theory.

Note that the normalization constant introduced in
(\ref{normalization}) involves the two quantities: intrinsic scale
$M_S$ and compactification radius $R$. These are related to each
other if it is assumed that the ``brane" model allows the graviton
to propagate in the full $(3+\delta)+1$ dimensional space-time.
Then the ordinary form of Newtons' gravitation law is only an
approximation valid at distances much greater than $R$. The
Newtonian gravitational constant (inverse square of the Planck mass
$M_P$) is obtained \cite{1} as a phenomenological parameter from
\begin{equation}
\left(\frac{M_P}{M_S}\right)^2=\left(2\pi M_S
R\right)^{\delta}=\frac{1}{\left(\rm Norm\right)^2} \ .
\label{Planck}
\end{equation}
Considering $M_P$ as an experimental input (and approximating
$R_1=R_2=\cdots =R_{\delta}$), shows via (\ref{Planck}) that $M_S$
is the only free parameter introduced to describe the extra
dimensional aspect of the present simple theory when $\delta$ is
fixed.

Next consider the Higgs ``potential" for the extended theory.
{}For simplicity we assume that the lepton conserving overlap term
$\Phi^{\ast}\Phi \phi^{\dagger}\phi$ is negligible. Then the Higgs
potential for the normal Higgs field $\phi$ is the same as in the
standard model while the Higgs potential for $\Phi$ is described
by the action
\begin{equation}
-\int d^4x d^{\delta}y\, \left[-c_0 M_S^{\delta +
2}\Phi^{\ast}\Phi+ c_1 M_S^{\delta} \left(\Phi^{\ast}\Phi\right)^2
\right] \ , \label{potential}
\end{equation}
where $\Phi$, as before, has engineering dimension equal to one.
The quantities $c_0$ and $c_1$ are positive (for spontaneous
breakdown of lepton number) and dimensionless. One might expect
$c_0$ and $c_1$ to be very roughly of order unity. The
minimization of (\ref{potential}) leads to the vacuum value
\begin{equation}
X\equiv \langle \Phi \rangle =\sqrt{\frac{c_0}{2c_1}}M_S \ .
\label{Phivev}
\end{equation}
We also find from (\ref{potential}) the increment to be added to
(\ref{increment}) for the ${\rm Re}\Phi_{n_1,n_2,\ldots}(x)$ fields:\
\begin{equation}
\Delta m^2_{n_1,n_2,\ldots}\left({\rm
Re}\Phi_{n_1,n_2,\ldots}\right) = 2 c_0 M^2_S \ .
\label{Reincrement}
\end{equation}
This result suggests that the ${\rm Re}\Phi_{n_1,n_2,\ldots}$
fields are too heavy to be produced by $Z$ decays. Finally,
consider the neutrino Yukawa terms in the extended theory. The
interactions of the usual Higgs field $\phi$ in (\ref{yukawa}) do not
change in the present model. However the lower right sub-block of
the matrix in (\ref{yukawa}) should now be gotten from the action
piece:
\begin{equation}
-\frac{1}{2}\int d^4x d^{\delta}y \, \rho^T_{a}(x)\,
\sigma_{2}\,\frac{{{\cal M}_{H}}_{ab}}{X} \,\rho_b(x)
\Phi^{\ast}(x,y)\delta^{\delta}(y)+{\rm h.c.} \ , \label{actionpiece}
\end{equation}
where ${{\cal M}_{H}}_{ab}/X$ is a matrix of Yukawa coupling
constants and the indices ($a,b$) go over only those for the heavy
singlet neutrinos ($4,5,6$). $\Phi(x,y)$ has the decomposition
\begin{equation}
\Phi(x,y)=X +
\frac{M_S}{M_P}\sum_{n_1,n_2,\ldots}\Phi_{n_1,n_2,\ldots}(x)
\exp\left[\frac{i}{R}\left(n_1 y_1+ n_2 y_2 + \cdots\right)\right]
\ , \label{decomposition}
\end{equation}
where (\ref{Planck}) was used. Substituting (\ref{decomposition}) into
(\ref{actionpiece}) shows, first, that the physical light neutrinos have
masses of the order (via the see-saw mechanism)
\begin{equation}
\frac{{\cal D}^2}{M_S\sqrt{\frac{c_0}{2c_1}} \left(\frac{{\cal
M}_H}{X}\right)} \ ,
\label{numassorder}
\end{equation}
where ${\cal M}_H/X$ is expected to be very roughly of the order
unity. Secondly, the Yukawa interactions of the Majoron and its
Kaluza-Klein excitations with the light neutrinos are described by
(c.f. (\ref{lj2})):
\begin{equation}
{\cal L}_{J}=\sum_{a=1}^3 \sum_{n_1,n_2,\ldots} \frac{i}{2X}
\frac{M_S}{M_P}J_{n_1,n_2,\ldots}m_a \left(\psi^T_a
C^{-1}\frac{1+\gamma_5}{2}\psi_a +
\bar{\psi}_a\frac{1-\gamma_5}{2}C \bar{\psi}^T_a\right) \ ,
\label{lj3}
\end{equation}
to leading order in the neutrino masses, $m_a$.
\section{$Z \rightarrow \nu \nu (J)$ decay}
This conceivable future test of a Majoron propagating in extra
dimensions is the focus of our present interest. Now we calculate
the amplitude for the $Z$ intermediate vector boson to decay into
two particular neutrinos and a
given member $ J_{n_1,...,n_{\delta}}$ of the Majoron Kaluza-Klein
tower. The Feynman diagram is shown in Fig. 1.
\begin{figure}[ht]
\centerline{\psfig{file=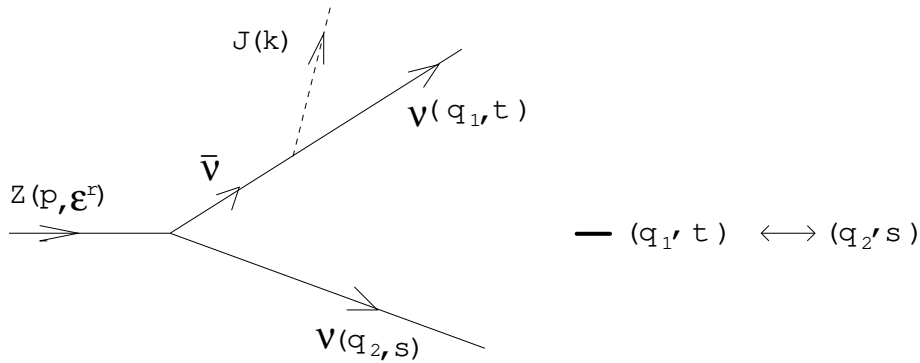,width=5in}}
\caption{Feynman diagram for $Z\rightarrow \nu \nu J$}
\end{figure}
The crucial coupling constant for the lepton number violating $(J)
\nu \nu$ vertex in the present model is read off from (\ref{lj3})
and (\ref{Phivev}) as
\begin{equation}
g_{a,b;n_1...n_{\delta}}=-\sqrt{\frac{2 c_1}{c_0}}
\:\frac{m_a}{M_p} \delta_{ab}, \label{KKg}
\end{equation}
correct to order $m_a$. Of course, our treatment could be applied to
any other Majoron model if $g$ in (\ref{KKg}) is appropriately modified. We
also need the usual $Z\bar{\nu} \nu$ coupling term of the standard model:
\begin{equation}
{\cal L}_{Z\bar{\nu}\nu}=\frac{-ie}{sin(2\theta_W)} Z_{\mu}
\sum_{a=1}^3 \bar \psi_a \gamma_{\mu} (\frac{1+\gamma_5}{2})
\psi_a , \label{Znunu}
\end{equation}
where e is the magnitude of the electron charge and $\theta_W$ is the
weak mixing angle.
 Then the desired amplitude is
\begin{eqnarray}
&&amp(Z\rightarrow \nu_a \nu_a J_{n_1,...,n_{\delta}})=
\frac{-e}{\sin(2\theta_W)}
g_{aa;n_1,...,n_{\delta}} \: {\varepsilon}^r_{\mu}(\vec p)\times
\nonumber\\
&&\bigg[\frac{1}{(q_1+k)^2} \bar{u}^s(\vec{q}_2) {\gamma}_{\mu}
(\frac{1+\gamma_5}{2}) {\gamma} \cdot (q_1+k) C \bar{u}^{tT}
(\vec{q}_1)-(q_1,t)\leftrightarrow (q_2,s)\bigg],\nonumber\\
&&
\label{amp}
\end{eqnarray}
correct to leading order in the neutrino mass $m_a$. Here
$\varepsilon_{\mu}^r(\vec p)$ is the Z polarization vector while
the other notations are shown in Fig. 1. Note that all the
neutrinos in (\ref{amp}) are being treated kinematically as two
component massless ones even though they should describe massive
Majorana fields. This is justified since
the coupling $g_{a,b;n_1...n_{\delta}}$ already contains a factor
$m_a$ and the corrections introduced by using the massive neutrino
propagator and spinors would be higher order in $m_a$ and hence
negligible.

We next take the squared magnitude of the amplitude (\ref{amp}) averaged
over Z polarizations. To leading order in $m_a$, only one polarization
state for each neutrino is needed in the calculation. After some
calculation, we find the following result, expressed in the Z rest frame:
\begin{equation}
\frac{1}{3}\sum_{Z pol.}\mid amp(Z\rightarrow \nu_a \nu_a
J_{n_1,...,n_{\delta}}) \mid^2 = \frac{e^2
g_{aa;n_1,...n_{\delta}}^2}{12 sin^2(2 \theta_W)} {\cal F}(E_1,E_2),
\label{sqamp}
\end{equation}
where
\begin{eqnarray}
{\cal F}(E_1,E_2)&=&\frac{E_1 E_2}{m_Z^2}\Bigg[
\frac{1}{(m_Z-2E_2)^2}
\bigg(8E_2(1-c)(E_2-m_Z)+2m_Z^2(3+c)\bigg)\nonumber\\
&+&\frac{1}{(m_Z-2E_1)^2}\bigg(8E_1(1-c)(E_1-m_Z)+2m_Z^2(3+c)\bigg)\nonumber\\
&-&\frac{2}{(m_Z-2E_1)(m_Z-2E_2)}
\bigg(8E_1E_2c(c-1)+4m_Z(E_1+E_2)(c-1)\nonumber\\
&+&2m_Z^2(3+c)\bigg)\Bigg]. \label{long}
\end{eqnarray}
In this formula, $E_1$ and $E_2$ are the energies of neutrino $1$ and
neutrino $2$, while $c$ is the cosine of the angle between their
momenta. One has
\begin{equation}
c=\frac{1}{2E_1E_2}[m_Z^2-m_J^2+2E_1E_2-2 m_Z(E_1+E_2)],
\label{cosine}
\end{equation}
where of course the neutrino masses have been taken to be zero.
${\cal F}(E_1,E_2)$ is seen to be invariant under the interchange
of $E_1$ and $E_2$. Finally the decay width, to particular
neutrinos and excited Majoron is given by:

\begin{equation}
\Gamma(Z \rightarrow \nu_a \nu_a J_{n_1...n_{\delta}})= \frac{e^2
g_{aa}^2} {48 (2\pi)^3 sin^2(2 \theta_W)}\frac{1}{m_Z} \int dE_1
dE_2 {\cal F}(E_1,E_2). \label{dw}
\end{equation}
To get the total contribution to the Z width we must sum over all
three neutrinos and over all kinematically allowed Majoron
excitations. In addition we should double the result for the inclusion
of $Z \rightarrow \bar{\nu} \bar{\nu} J.$
\section{$Z \rightarrow \nu \nu J$ decay in $3+1$ Majoron models}
The partial width for $Z \rightarrow \nu \nu J$ in ``usual'' 3+1
dimensional Majoron models is expected to be small and, probably for
this reason, does not seem to have been previously treated. Thus it is
of some interest to present this case first. We will also see that it
provides a useful ``warm up'' for the higher dimensional situation.

In the present case we can use the formulas of the last section and
just disregard the excited Majoron states. There is only a zero mass
Majoron. Our job is to numerically integrate (\ref{dw}) with integrand
(\ref{long}). In addition we should set $m_J$ to zero in (\ref{cosine}).
Incidentally, the phase space boundary for integration is gotten
by setting $\mid c \mid=1$. The resulting boundary is shown in Fig. 2.
\begin{figure}[ht]
\centerline{\psfig{file=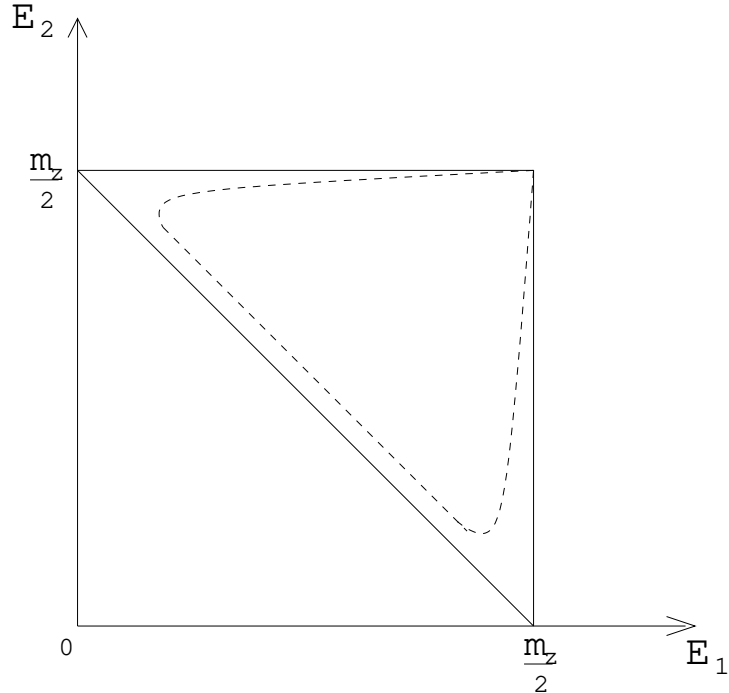,width=4in}}
\caption{Phase space boundary for $m_J=0$. The solid triangular region
shows the $m_{\nu}=0$ boundary, while the dashed curve is a rough
sketch of the $m_{\nu}\neq 0$ case}
\end{figure}
The simple triangular region corresponds to the plausible
kinematic approximation of zero neutrino masses. Now we can see
that there is a problem with this approximation; the denominators
in (\ref{long})
vanish on the boundary lines $E_1=m_Z/2$, $E_2=m_Z/2$. To handle
this we solve for the phase space boundary including the effect of
non-zero neutrino mass. Generalizing (\ref{cosine}) and setting
the new
 $\mid c\mid $ equal to unity yields, to leading order in $m_{\nu}$, the three
components of the boundary curve:
\begin{eqnarray}
E_2&=&\frac{m_Z}{2}-m_{\nu}^2 \bigg(\frac{m_Z-2 E_1}{4E_1
m_Z}\bigg)+\cdots,\nonumber\\ E_1+E_2&=& \frac{m_Z}{2}+m_{\nu}^2
\bigg(\frac{m_Z}{4 E_1(m_Z-2 E_1)}\bigg)+{\cdots},\nonumber\\
E_1&=&\frac{m_Z}{2}-m_{\nu}^2 \bigg(\frac{m_Z-2 E_2}{4E_2
m_Z}\bigg)+{\cdots}. \label{boundary}
\end{eqnarray}
This boundary, which is symmetric about the line $E_1=E_2$, is
sketched in Fig.2, wherein the deviation from the triangular
boundary is greatly exaggerated for clarity. Only at the single
point $E_1=E_2=m_Z/2$ does the true boundary coincide with the
triangular boundary. Except for this point there is no possibility
of a divergence. Physically, this point corresponds to the Majoron
carrying vanishing energy and the two neutrinos coming off back to
back.

To go further, we simplify the expression obtained by substituting
(\ref{cosine}) into (\ref{long}) with $m_J=0$, to find:
\begin{equation}
{\cal F}(E_1,E_2)\arrowvert_{m_J=0}  =
2\bigg(\frac{E_1-m_Z/2}{E_2-m_Z/2}+\frac{E_2-m_Z/2}{E_1-m_Z/2}\bigg)+8
\bigg(\frac{E_1}{m_Z}+ \frac{E_2}{m_Z}-1\bigg).
\label{exactsquamp}
\end{equation}
{}From this expression we see that ${\cal F}$ can be consistently
defined to be finite at $E_1=E_2=m_Z/2$ so there is really no
divergence anywhere in the physical region. Because of its simple
form it is straightforward to analytically integrate
(\ref{exactsquamp}) within the approximate boundary
(\ref{boundary}) to obtain the partial $Z$ width for a particular$
J \nu_a \nu_a$ final state:
\begin{equation}
\Gamma(Z \rightarrow \nu_a \nu_a J) \approx \frac{e^2 g_{aa}^2
m_Z} {48 (2\pi)^3 sin^2(2 \theta_W)}
\left[\ln\left(\frac{m_Z}{m_a}\right)-\frac{4}{3}\right].
\label{log1}
\end{equation}
For sensible values of the neutrino masses  $m_a\leq 1 \: eV$, the $\ln$
term is the dominant one. We see that the rate vanishes as
$m_a\rightarrow 0$ since (\ref{gexp}) shows that $g_{aa}^2=m_a^2/X^2$ and
this factor overcomes the potentially troublesome $\ln (m_a)$. In the
Majoron model discussed in section 2, $m_a=0$ corresponds to a phase
of the theory in which lepton number in not spontanously broken and
the field $J= \textrm {I}m \Phi$ is not massless.

Using the well known experimental numbers for $e^2$, $m_Z$ and
$\theta_W$ in (\ref{log1}) together with a choice $m_a\approx 1
eV$ for all three $\nu_a$'s and all three $\bar{\nu_a}$'s gives
the following estimate for the partial width associated with
$Z\rightarrow J +2\: \textrm {neutrinos}$:
\begin{equation}
\Gamma(Z\rightarrow J+ 2 \:\textrm {neutrino}s)\approx 0.14 \: g_{aa}^2 \quad
GeV.
\label{estimate}
\end{equation}
Comparing this with the uncertainty in the invisible width of the $Z$ quoted
in (\ref{invisiblewidth}) gives a rough bound on the coupling constant for
the Majoron
with two neutrinos:
\begin{equation}
\mid g_{aa}\mid \leq 0.11.
\label{rbound}
\end{equation}
Here, of course, the assumption has been made that known decays can
accurately account for the central value. This assumption, as well as the
uncertainty itself, should improve in the future. As expressed in
(\ref{rbound}), the experimental bound on $\mid g_{aa} \mid$ can be applied to a
more general Majoron theory (e.g., one with an arbitrarily more
complicated Higgs sector) than the one given in section 2). In the
present case, the bound (\ref{rbound}) is very weak since $g_{aa}=\frac{m_a}{X}
\approx 10^{-13} $, if X is generously chosen to be only as large as
$10^4 \,GeV$. In the extra dimensional version of the Majoron theory
the invisible width is expected to be greatly increased due to the large number
of extra channels with excited Majorons $J_{n_1,n_2,...,n_{\delta}}$. This might be
expected to greatly strengthen the bound in certain scenarios.

\section{$Z \rightarrow \nu \nu (J)$ decay in $(3+\delta)+1$ Majoron models}
Now it is necessary to consider the decay to each
$J_{n_1,...,n_{\delta}}$ and then sum them all up. The rate for each
separate mode is given by (\ref{dw}) wherein the $m_J^2$ term in (\ref{cosine}) is no
longer neglected. The phase space boundary curve which replaces (\ref{boundary})
now has the components:
\begin{eqnarray}
E_2=\frac{m_Z}{2}-\frac{m_J^2} {2m_Z-4E_1}\ , \qquad E_1+E_2=
\frac{m_Z^2-m_J^2}{2m_Z}\ . \label{boundary2}
\end{eqnarray}
Here the neutrino masses have been neglected. It is easy to see
that the lines $E_1=m_Z/2$ and $E_2=m_Z/2$, where the expression
(\ref{long}) blows up, lie outside the present phase space
boundary so there is no question of a divergence. The situation is
formally similar to the massive photon method of regulating the
infra red divergent diagrams in QED. However the non-zero mass for
$J$ is not an artifice here. For the case of the ground state
$J_{0,...,0}$ of zero mass, the neutrino mass must strictly
speaking be restored as in the last section.

We have carried out a numerical integration of the factor $\int
dE_1 dE_2 {\cal F}(E_1,E_2)$ in (\ref{dw}) for relevant values of
$m_J$ with the phase space boundary of (\ref{boundary2}) above.
The results are graphed in Fig. 3. It is seen that for smaller
values of $m_J$ (less than about $10$ GeV) the integral is a
straight line function of $\ln(m_J)$, while for large $m_J$ it is
very small, quickly vanishing as $m_J \rightarrow m_Z$.
\begin{figure}[htb]
\centerline{\psfig{file=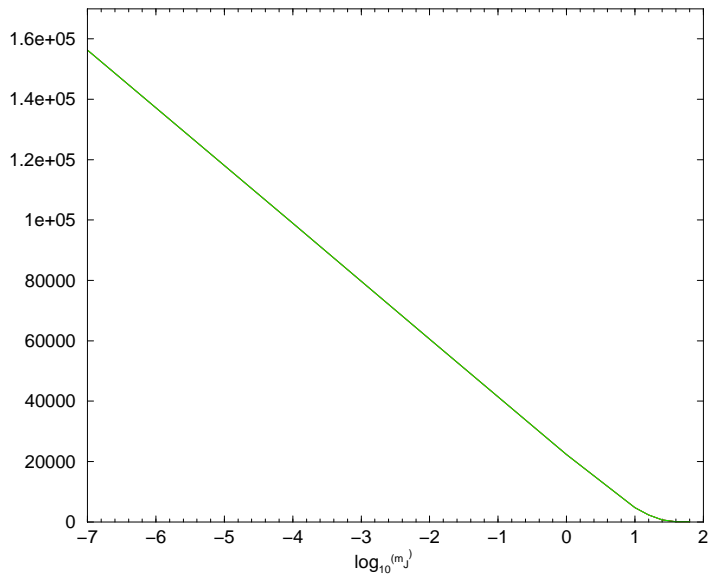,width=3in,angle=270}}
\caption{Plot of the integral in (\ref{dw}) vs $ Log_{10}(m_J \textrm {in GeV})$}
\end{figure}

The small $(m_J/m_Z)$ dependence of the integral may be understood
in the following way. On substituting (\ref{cosine}) into
(\ref{long}) we find that the effect of $m_J\neq   0$ is simply to
add to the $m_J=0$ expression (\ref{exactsquamp}), pieces which
vanish at least as fast as $m_J^2$. Thus for small $m_J$ it should
be a good approximation to still use (\ref{exactsquamp}) for the
integrand. Then, analogously to the previous section, it is
straightforward to analytically integrate (\ref{exactsquamp}) over
the phase space boundary (\ref{boundary2}) (with lower limits of
the integration energies around $m_{\nu}\approx 1 eV$). This
yields for small $m_J/m_Z$;
\begin{equation}
\int dE_1 dE_2 {\cal F} (E_1,E_2) \approx m_Z^2 \left[
\ln\left(\frac{m_Z}{m_J}\right)-{\cal O}(1)\right] \label{log2}
\end{equation}
where the ln term is dominant. Comparison with (\ref{log1}),
obtained with the same integral but with the different boundary
(\ref{boundary}), is instructive. Most of the contribution to the
integral comes from $E_1$ and $E_2$ close to the singularities
$E_1=m_Z/2$ and $E_2=m_Z/2$ lying just outside the physical phase
space boundary. In the case of (\ref{log1}), $m_{\nu}$ specifies
the closeness of the unphysical singularity while $m_J$ takes over
this role in the case of (\ref{log2}).

To go further it is convenient to make a fairly realistic analytic
approximation to the ``exact'' numerical result illustrated in
Fig. 3. Since the rate integral is very small beyond $m_J \approx 10$
GeV the simplest procedure is to  just set it to zero beyond a certain point
and use the form (\ref{log2}) for the main region:
\begin{equation}
\int dE_1 dE_2 {\cal F}(E_1,E_2) \approx  \left\{
\begin{array}{ll} m_Z^2 [\ln(\frac{m_Z}{m_J})-A]& \textrm{if
$\ln(\frac{m_Z}{m_J})>A$}\\ 0&\textrm{if $\ln(\frac{m_Z}{m_J})<A$}
\end{array} \right .
\label{approx}
\end{equation}
The parameter choice $A \approx 1.75$, which corresponds to a cut
off of $m_J \approx 15.8$ GeV, gives a good fit to the main low
$m_J$ region.

We next must sum over the rates for each $m_J$. The members of the
Majoron tower are labeled by integers $ n_1,...,n_{\delta}$ and
have squared masses given by (\ref{increment}). In the interesting
case of relatively low intrinsic scale, $M_S$ there are a huge
number of them. For example, using (\ref{Planck}) to relate $R$ to
$M_S$, shows that with $M_S=10^4$ GeV and $\delta=1$, the first
Majoron excitation will have a mass of order $10^{-26}$ GeV. Thus
we can safely replace summation by integration. Denoting
$\Gamma_Z(m_J)$ as the $Z$ width to two neutrinos plus a Majoron
excitation, the width for decay to the whole Majoron tower is then

\begin{equation}
\Gamma(Z\rightarrow \nu \nu (J)+\bar\nu \bar\nu(J))= \frac{2
\pi^{\delta/2}}{\Gamma(\delta/2)} R^{\delta} \int_0^{m_Z}
m_J^{\delta-1} \Gamma_Z(m_J) dm_J, \label{tower}
\end{equation}
where the prefactor is associated with the ``area'' of a unit
hypersphere in the $\delta$ dimensional space needed to count the
degenerate modes in (\ref{increment}). $\Gamma(\delta/2)$ is the
gamma function. With the approximation (\ref{approx}) it is simple
to do the integration in (\ref{tower}) analytically and get the
final result:

\begin{eqnarray}
\Gamma(Z\rightarrow \nu \nu (J)+\bar\nu \bar\nu(J))&\approx &
\frac{e^2 g_{aa}^2 m_Z}{64 \pi^3 \sin^2(2 \theta_W)} \left[\frac{2
\pi^{\delta/2}}{\Gamma(\delta/2) (2 \pi)^{\delta}}\right]
\left[\frac{M_P}{M_S}\right]^2 \, \frac{1}{\delta^2}
\left(\frac{m_Z}{M_S} e^{-a}\right)^{\delta}. \label{result}
\end{eqnarray}
In obtaining this form we used (\ref{dw}) and also (\ref{Planck})
(for eliminating $R$ in favor of $M_S$ and $\delta$). Eq.
(\ref{result}) can quickly give an idea of how the rate depends on
the choice of $\delta$, the number \footnote{For larger values of
$\delta$ the approximation based on (\ref{approx}) gets worse
since higher moments of $\Gamma_Z(m_J)$ are required.} of extra
dimensions, and $M_S$, the intrinsic mass scale. Note that $
A\approx 1.75$ is not an arbitrary parameter but is associated
with the approximation to the exact numerical integration given in
(\ref{approx}). $g_{aa}$, the coupling constant of each member of
the Majoron tower to two neutrinos, will at first be regarded as a
quantity to be bounded by comparison with experiment.

Comparing (\ref{result}) with the result (\ref{log1}) for the Majoron
model in $3+1$
dimensional space-time shows that, for the \underline{same} coupling
constant $g_{aa}$, there may be a big amplification factor
 $(\frac{M_P}{M_S})^2$. If $M_S$ is chosen to be $10^4 GeV$,
corresponding to the range which should be probed in the next
generation of accelerators, this amplification factor is about
$10^{30}$. It is due to the large value of the compactification
radius
which results in a very large number of closely spaced states in
the Kaluza Klein tower. For example with $M_S=10^4 GeV$ and
$\delta=1,2,3,4$ we have respectively $R=2.37\times 10^{25}
GeV^{-1} (4.58 \times 10^9 m), 1.94\times 10^{10} GeV^{-1}
(3.82\times 10^{-6} m), 1.3\times 10^{5} GeV^{-1} (3.57 \time
10^{-11} m), 5.57 \times 10^{2} GeV^{-1} (1.10 \times 10^{-13}
m)$. The first excited Majoron has a mass, from (\ref{increment}),
$R^{-1}$. It is well known that $\delta \approx 1$ is ruled out
for a model of this type since $R$ is evidently large enough to
contradict Newton's gravtational force law. Eq(\ref{result}) shows
that, when $M_S$ is fixed, the main dependence on $\delta$ is due
to the factor $(m_Z/M_S)^{\delta}$.

Substituting numbers into (\ref{result}) gives for
$\delta=1,2,3,4$ the respective predicted widths (in $GeV$)
$4.53\times 10^{24} g_{aa}^2$,
 $8.97\times 10^{20} g_{aa}^2$, $2.01\times 10^{17} g_{aa}^2$,
 $4.48\times 10^{13} g_{aa}^2$. If the uncertainty of the $Z$'s
invisible width $1.7 \times 10^{-3} GeV$ is roughly taken as an
indication of the maximum allowed value for the total width into a
Majoron and two neutrinos these numbers can be interpreted as the
following bounds on $\mid g_{aa} \mid$: $\mid g_{aa} \mid <1.9
\times 10^{-14}$, $1.4 \times 10^{-12}$, $9.2 \times10^{-11}$,
$6.2\times10^{-9}$ for $\delta=1,2,3,4$ respectively. These are
much stronger bounds than the one obtained in (\ref{rbound}) for
the model in $3+1$ space-time dimensions.

We still must ask what is the natural value of $g_{aa}$ in the
simple model presented in section II. According to (\ref{KKg}) and
the discussion of section II we would expect the very small value
$\mid g_{aa} \mid \sim m_{\nu}/M_P \approx 10^{-28}$. Then the
predicted width would be (for $\delta=2$) of the order $10^{-36}
GeV$. This is even smaller than the order $10^{-27} GeV$ expected
from (\ref{estimate}) in the $3+1$ dimensional case. What is
happening is that the $(\frac{M_P}{M_S})^2$ enhancement in
(\ref{result}) is being cancelled by a suppression
factor$(\frac{M_S}{M_P})^2$ in $g_{aa}^2$. Also the last factor in
(\ref{result}) provides additional suppression. Of course these
results can be modified if we are willing to accept (technically
unnatural) fine tuning of the parameters. In (\ref{KKg}) we would
need to fine tune $c_1/c_0$ to be extremely large; this
corresponds to an exceptionally small wrong sign mass squared term
in the Higgs potential (\ref{potential}). In addition the Yukawa
coupling (${\cal M}_H/X$) which appears in (\ref{potential}) would
have to be fine tuned very large in order to keep the neutrino
masses of correct order. It is possible that this fine tuning
could be made natural in a supersymmetric version of the Majoron
theory or with a special dynamical mechanism. Furthermore the
singlet Majoron model in section 2 is the simplest one. Enlarging
it by including more Higgs fields should also modify the coupling
constants. Thus it is conceivable that extra dimensional theories
could lead to enhancement of the $Z$'s partial width for decay to
a Majoron tower and two neutrinos.

\section{Summary}

We investigated the possibility that the accurately known value of
the Z width might be used to get information about the process Z
$\rightarrow$ J $+$ two neutrinos. Here J is a Majoron-- the
Goldstone boson associated with a proposed mechanism for
generation of neutrino mass by spontaneous breakdown of lepton
number. It was noted that the main contribution to the process
comes from the kinematical region near the phase space boundary,
outside of which the matrix element is singular. This led to a
simple analytic form for the partial width. A bound on the
dimensionless lepton number violating coupling constant $g_{aa}$
of the Majoron to two neutrinos was estimated to be $\mid g_{aa}
\mid \leq 0.11$. However in the simple singlet Majoron theory
discussed in section II, the expected magnitude is more like $\mid
g_{aa} \mid ={\cal O}(10^{-13})$. Thus the bound is not very
restrictive although it is possible that more complicated Majoron
models might predict larger values for $\mid g_{aa} \mid$ .

The treatment above was generalized to the case where Physics is
described by a "brane" embedded in a space of $\delta$ extra
dimensions, all toroidally compactified to radius R. In this case
there are typically a very large number of excited Majorons. A
simple approximate formula was derived for the width to all of the
J's plus two neutrinos for any value of $\delta$ and the intrinsic
scale $M_S$. In the case $\delta =2$ and $M_S = 10^4$ GeV the
bound is estimated as $\mid g_{aa} \mid \leq 1.4 \times 10^{-12}$,
which appears considerably stronger than the ordinary one. However
the coupling of ``brane" particles to ones like the J, which can
propagate in the extra dimensions, is greatly suppressed. Thus in
an extra dimensional Majoron theory without special fine tuning of
the parameters, the expected value of $\mid g_{aa} \mid$ is only
about $10^{-28}$. The net effect of introducing extra dimensions
is a suppression, rather than an enhancement, of the decay rate
into two neutrinos plus a Majoron tower. If one relaxes the
prescription of ``no fine tuning" it is possible to obtain an
enhancement. The same may be conjectured for possible alternative
Majoron schemes in extra dimensions.

\acknowledgments The work of S.N and J.S. has been supported in
part by the US DOE under contract DE-FG-02-85ER 40231 while the
work of F.S. has been partially supported by the EU Commission
under contract HPRN-CT-2000-00130. F.S. thanks Y. Takanishi for
discussions. One of the authors (S. Moussa) would like to thank
the Egyptian Ministry of Higher Education for support.


\begin{thebibliography}{9}

\bibitem{1} N. Arkani-Hamed, S. Dimopoulos and G. Dvali, Phys. Lett. {\bf
B429},263(1998); Phys. Rev. {\bf D59},086004(1999).

\bibitem{RS} L. Randall and R. Sundrum, Phys. Rev. Lett. {\bf 83},
3370(1999).

\bibitem{DDG} K. R. Dienes, E. Dudas and T. Gherghetta, Nucl. Phys.{\bf
B557},25(1999).

\bibitem{ADDM} N. Arkani-Hamed, S. Dimopoulos, G. Dvali and
J. March-Russell, hep-ph/9811448.

\bibitem{DS} G. Dvali and A. Y. Smirnov, Nucl. Phys. {\bf B563},63(1999).

\bibitem{BGS} R. Barbieri, P. Geminelli and A. Strumia, Nucl. Phys. {\bf
B585},28(2000).

\bibitem{IV} A. Ionnissian and J. W. F. Valle, Phys. Rev. {\bf
D63},073002(2001).

\bibitem{FP} A. E. Faraggi and M. Pospelov, Phys. Lett. {\bf
B458},237(1999).

\bibitem{MN} G. C. McLaughlin and J. N. Ng, Phys. Lett. {\bf
B470},157(1999).

\bibitem{DK} A. Das and O. Kong, Phys. Lett. {\bf B470},149(1999).

\bibitem{C} C. D. Carone, Phys. Rev. {\bf D61},015008(2000).

\bibitem{BDN} T. Banks, M. Dine and A. E. Nelson, JHEP 9906: 014(1999).

\bibitem{MPP} R. N. Mohapatra, A. Perez-Lorenzano and C. A. de S. Pires,
Phys. Lett. {\bf B491},143(2000).

\bibitem{MRS} E. Ma, M. Raidal and U. Sarkar, Phys. Rev. Lett. {\bf
85},3769(2000).


\bibitem{rpp} Review of Particle Physics. D.E. Groom et al. Eur. Phys. J. {\bf C15},1(2000).

\bibitem{TPV} R. Tomas, H. Pas and J. W. F. Valle, hep-ph/0103017.

\bibitem{CMP} Y. Chikashige, R. N. Mohapatra and R. D. Peccei, Phys. Lett.
{\bf B98},265(1981).

\bibitem{SV} J.~Schechter and J.W.F.~Valle, Phys.~Rev.~{\bf D25}, 774
(1982).

\bibitem{ss} T. Yanagida, Proc. of the Workshop on Unified Theory and
Baryon Number in the Universe, ed. by O. Sawada and A. Sugamato (KEK
Report 79-18,1979), p 95;
M. Gell-Mann, P. Ramond and R. Slansky in Supergravity, eds P. van
Niewenhuizen and D. Z. Freedman (North Holland, 1979);
 R. N. Mohapatra and G. Senjanovic,
Phys. Rev. Lett. \textbf{44}, 912 (1980).

\end{thebibliography}
\end{document}